# R-graphyne: a new two-dimension carbon allotrope with versatile Dirac-like point in nanoribbons

Wen-Jin Yin,[1,2] Yue-E Xie,*[1] Li-Min Liu,*[2] Ru-Zhi Wang,[3,2] Xiao-Lin Wei[1,2], and Leo Lau,[4,2] Yuan-Ping Chen,[1]

[1]Department of Physics and Laboratory for Quantum Engineering and Micro-Nano Energy Technology, Xiangtan University, Xiangtan 411105, Hunan, China

[2]Beijing Computational Science Research Center, Beijing 100084, China

[3]Laboratory of Thin Film Materials, College of Materials Science and Engineering, Beijing University of Technology, Beijing 100124, China

[4]Chengdu Green Energy and Green Manufacturing Technology R&D Center, Chengdu, Sichuan, 610207, China

*Corresponding author: limin.liu@csrc.ac.cn，xieyech@xtu.edu.cn

**Abstract**

A novel two-dimension carbon allotrope, rectangular graphyne (R-graphyne) with tetra-rings and acetylenic linkages, is proposed by first-principles calculations. Although the bulk R-graphyne exhibits metallic property, the nanoribbons of R-graphyne show distinct electronic structures from the bulk. The most intriguing feature is that band gaps of R-graphene nanoribbons oscillate between semiconductor and metal as a function of width. Particularly, the zigzag edge nanoribbons with half-integer repeating unit cell exhibits unexpected Dirac-like fermions in the band structures. The Dirac-like fermions of the R-graphyne nanoribbons originate from the central symmetry and two sub-lattices. The extraordinary properties of R-graphene nanoribbons greatly expand our understanding on the origin of Dirac-like point. Such findings uncover a novel fascinating property of nanoribbons, which may have broad potential applications for carbon-based nano-size electronic devices.

## I. Introduction

Graphene, as a typical allotrope of carbon, has attracted great attentions from different fields because of the two-dimensional (2-D) $sp^2$ bond structure with unique electronic properties, for instance, half-integer quantum Hall effect, high migration rate, and mass less carriers.[1, 2, 4] The origin of such amazing electronic properties of graphene is attributed to its peculiar band structure featuring so-called Dirac points.[2] In the vicinity of the Dirac point, the valence band and conduction one meet at the Fermi level forming a double cone, which appears in a linear relationship. The emergence of Dirac fermions originate from $\pi$ and $\pi^*$ bands and the hexagonal honeycomb. The unique properties of the Dirac fermions suggest the graphene to have potential to be as a revolutionary material for the future generation of high-speed nano-electronics.[5-7] Many attentions have been attracted to explore the extraordinary electronic properties.

Recently, many new materials, such as the new carbon allotropes, the BN-doping graphene, topological insulator, have been proposed.[8-11, 13-16] Graphyne can form various types by inserting carbon triple bonds (C≡C) into C-C bonds in graphene, and three highly symmetric forms are named as $\alpha$-graphyne, $\beta$-graphyne, and 6,6,12-graphyne, respectively.[11, 15] Recent study suggested that the three graphyne of carbon allotropes exhibit Dirac feature in the electronic properties similar to graphene, and 6,6,12-graphyne has even better electronic properties than graphene.[11, 17] Different from graphene, 6,6,12-graphyne does not have hexagonal symmetry, and graphyne exhibits two self-doped nonequivalent distorted Dirac cones, suggesting that electronic properties may be even more amazing than graphene.[17] Recently, several carbon allotropes have been proposed, which do not own hexagonal symmetry. For instance, T-carbon, which contains sub-lattices, also own Dirac cones as graphene and graphyne.[9, 12, 13]

In the real application, either graphene or other carbon allotropes should be used as nanoribbons with edges. Although the graphene own the property of the Dirac clone, such feature generally disappears in the nanoribbons. For example, the zigzag-edged graphene is metallic with peculiar edge states on both sides of the nanoribbons regardless of its widths, while the armchair-edged graphene can be either metallic or semiconducting depending on their width arising from both quantum confinement and edge effect.[18-21] Similar to graphene, recent studies show that all the tailored graphyne nanoribbons are semiconductor, and the band gaps of the graphyne nanoribbons decrease with the increase of widths.[22] Although quite a lot bulk materials owns Dirac-like fermions, but such important feature generally disappears for the nanoribbons. The lost of the electronic properties in nano-size may greatly limit the application of graphene in high-speed electronic devices.[23, 24] To fully utilize the electronic properties of Dirac-like in the nano-device, it requires that the nanoribbons posses Dirac-like points in the band structure as in the 2-D ones. Considering most of current materials only show the Dirac-like properties in 2-D, it is necessary to explore a new material that exhibits the electronic properties of Dirac clone in 1-D nanoribbon.

In this paper, a new carbon allotrope rectangular graphyne (R-graphyne), which consists of tetra-rings and acetylenic linkages, is explored by first principles calculations. R-graphyne is thermodynamically stable, compared with other carbon allotropes. Although 2-D bulk R-graphyne is metallic as the valence band cross the Fermi level, the zigzag R-graphyne nanoribbons exhibit half-integer oscillating between metallic and semiconductor. The most amazing finding is that the half-integer widths zigzag R-graphyne nanoribbons show Dirac-like points in the band structures. As for the integer width, the zigzag nanoribbons appear a direct band gap (~0.25 eV). The detailed analysis of electronic properties reveals that the Dirac-like fermions in the zigzag nanoribbons origin from the two sub-lattices, which is orientation-dependent. Meanwhile, the origination of the Dirac-like point is in relation with the central symmetry instead of the rotation symmetry of the charge density distribution of the nanoribbons. The results suggest a novel carbon allotrope with Dirac-like point of intriguing property for 1-D nanoribbon, and such results greatly expand understanding of the Dirac point in 1-D. The new R-graphyne in 1-D nanoribbons has many potential applications for high-performance electronic materials.

## II. Computational method

The first-principles calculations based on the density functional theory (DFT) have been carried out by means of the Vienna ab initio simulation program package (VASP).[22] Perdew-Burke-Ernzerhof (PBE) functional is employed for the exchange-correlation term according to generalized gradient approximation (GGA).[23] The projector augmented wave (PAW) method has been used to represent the atomic cores. The energy cutoff is set at 500 eV for the plane-wave basis, which yields total energies converged at least 0.001 eV per atom, as confirmed by calculations with higher cutoffs.[25] The convergence criterion of the self-consistency process is set to $10^{-7}$ eV between two ionic steps. A Monkhorst-Pack **k**-point grid with 11×11 **k** points in the xy plane and one **k** point in the $z$ direction is chosen in some cases including the Γ-point. A vacuum region of 15 Å is added to avoid interactions between adjacent images. The geometries are fully relaxed until the residual force is less than 0.001 eV/Å on each atom. The stability of the configuration is calculated by phonon calculations, which are performed with the Phonopy package[26] along with the VASP. To further examine the thermodynamics stability of the R-graphyne, the molecular dynamics simulation is performed with Reactive Force Field force field (RexFF), as implanted in LAMMPS.[27,28]

## III. Results and Discussion

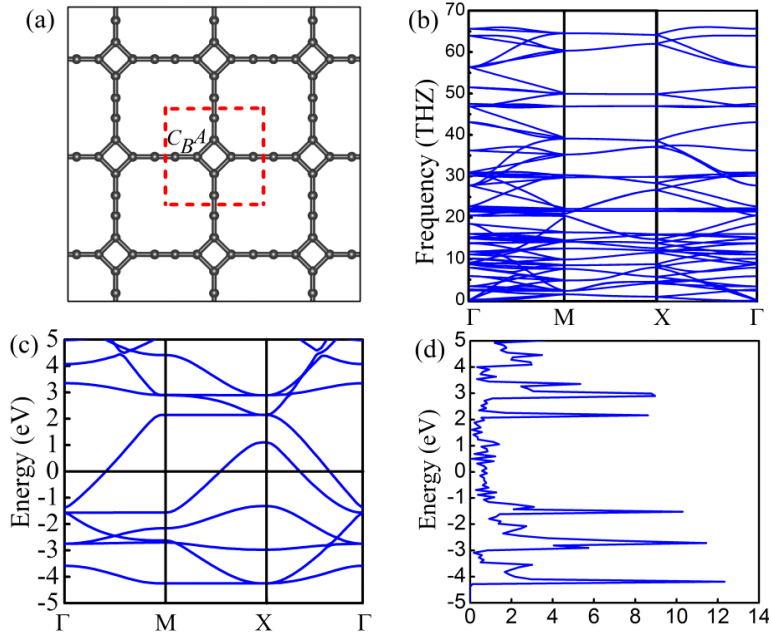

Fig. 1 (color online) (a) Geometrical structure of the R-graphyne and three types of bonds are denoted as *A*, *B* and *C*, respectively. (b) The phonon spectra of the R-graphyne. (c) The electronic structure and

(d) the total density of the 2-D R-graphyne sheet. The locations of points are Γ (0, 0), M (0.5, 0.5), and X (0, 0.5) are appropriate for both the phonon and band structure. The Fermi energy is set to 0 eV, similarly hereinafter. The black balls represent carbon atoms.

A new carbon allotrope of 2-D plane R-graphyne sheet is firstly explored, as shown in Fig. 1(a). The R-graphyne contains tetra–rings (C4) in sp$^2$ bond and the acetylenic linkages (C≡C) in sp bond. The structure can be derived from inserting the carbon triple bond (C≡C) into one thirds of C-C in the plane T-graphene[9] between two neighboring units or transformed the hexagonal into tetra rings of graphyne. According to the bond length, the bonds can be classified into three types as *A*, *B*, and *C*, where *A*, *B*, and *C* denote the bond length of the C-C in C4, C-C between C4 and C≡C, and C≡C, respectively. The sheet of R-graphene can be described by the plane group *p4mm* with lattice constant of *a*=*b*=8.52 Å. The R-graphyne of the primitive cell contains eight atoms, as shown in red dashed line in Fig. 1(a). The bonds length of *A*, *B*, and *C* are 1.470 Å, 1.350 Å, and 1.246 Å, respectively. It should be noted that the bond length of *A/B* is longer/shorter than that of C-C in graphene, respectively. The bond length of *C* is close to that of the acetylenic linkage.

In order to know the stability of the R-graphyne, the formation energies are calculated along with the other carbon allotropes.[12, 30-31] Compared graphene (-9.26 eV/atom), buckled T-graphene (-8.41 eV/Å)[9], bcc-C$_8$ (-8.49 eV/atom)[30], and graphdiyne (-8.49 eV/atom)[10], R-graphyne has a comparable formation energy of -8.40 eV/atom, which is rather close to that of T-graphene or graphdiyne. To further confirm the thermodynamic stability, the phonon vibration frequencies are calculated, and the result is shown in Fig. 1(b). It can be seen that all branches of the phonon spectrum are positive and no any imaginary phonon mode exists, suggesting the stability of R-graphyne. It is instructive to note that the highest phonon frequency of R-graphyne is estimated to be 66 THz, very close to that of graphyne (67 THZ).[25] To further check the stability of R-graphyne, a molecular dynamics simulation at temperature of 500 K with a time step of 1 fs is carried out, and the geometry of R-graphyne is retained for 5 ps, which also suggests that the R-graphyne is thermodynamically stable[28].

The calculated electronic properties and total density of states (DOS) of R-graphyne are depicted in Fig. 1(c) and Fig. 1(d), respectively. It is obvious that the valence band of R-graphyne passes through the Fermi level (E$_F$), indicating that R-graphyne is metallic. The DOS around the E$_F$ is about 0.09 states/eV/atoms. Previous studies showed that the (4, 4) carbon nanotube has the largest DOS of 0.07 states/eV/atom at E$_F$ among all plausible metallic carbon nanotube.[31] Interestingly, R-graphyne sheet has an even larger DOS than (4, 4) carbon nanotube at E$_F$, suggesting great potential for nano-electronic. Compared with graphyne, a direct band gap appears at M point originating from the p$_z$ orbit in the hexagonal rings.[10] However, different from the sp$^2$ bond of graphene, some sp$^2$ bonds of R-graphyne are not entirely standard bond, and the hexagonal symmetry of R-graphyne disappears in forming tetra-ring. As a result, the $\pi$-$\pi^*$ gap located at M point near the E$_F$ of graphyne is filled in R-graphyne (see Fig. 1(d)), which is similar to DOS of

T-graphene.[9]

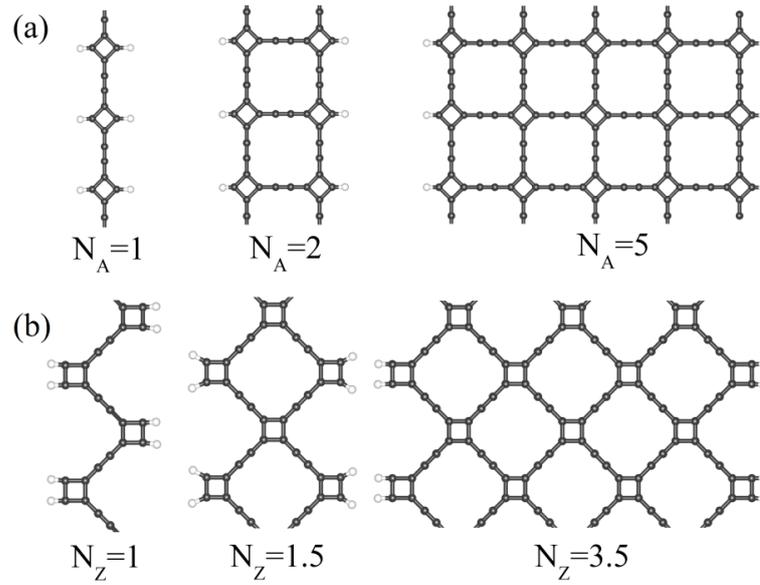

Fig. 2 (color online) The configurations of R-graphyne nanoribbons obtained by cutting through an infinite R-graphyne along two directions. (a) Armchair nanoribbons of R-graphyne with widths $N_A$=1, 2 and 5, where $N_A$ denotes the number of chains of tetra carbon rings. (b) Zigzag edged R-graphyne nanoribbons with widths $N_Z$=1, 1.5, and 3.5 expressed by $N_Z$. The black and white balls represent carbon and hydrogen atoms, respectively.

Structure tailoring can induce many new properties such as in graphene and graphyne.[21, 22] Similar to the graphene and graphyne nanoribbons, R-graphyne sheet can also be patterned into two typical styles of nanoribbons along different orientations, namely, armchair and zigzag R-graphyne nanoribbons. The atomic structure of armchair and zigzag R-graphyne nanoribbons are shown in Fig. 2(a) and 2(b), respectively. The width of armchair nanoribbon is denoted as $N_A$ to distinguish the number of the repeating units. The widths $N_A$ from 1 to 5 are considered in our calculations. Different from the armchair nanoribbons, the number of repeating units for the zigzag R-graphyne nanoribbons can be classified into two types by a half-integer width of repeating cell, which is denoted as $N_Z$ in Fig. 2(b) the stripe widths from $N_Z$=1 to $N_Z$=5.5 are used to examine the properties of zigzag R-graphyne. During the geometric optimization, all the edge carbon atoms are saturated with hydrogen atoms.

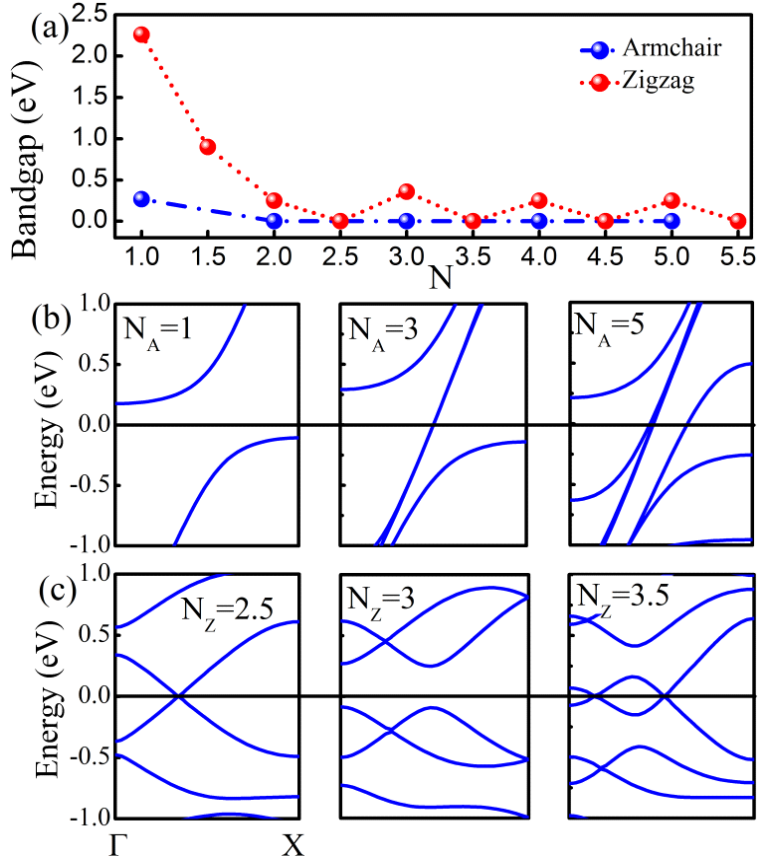

Fig. 3 (color online) The calculated electronic properties of the R-graphyne. (a) The band gaps of R-graphyne nanoribbons with various widths. The band structures of (b) armchair edged R-graphyne with $N_A$=1, 3 and 5; (c) zigzag edged R-graphyne with $N_Z$=2.5, 3 and 3.5.

As mentioned above, the bulk R-graphyne is metallic with high states at $E_F$. Hence, it is natural to ask whether nanoribbons of R-graphyne also have the same kind of electronic properties. The calculated band gaps of nanoribbons as a function of the widths are displayed in Fig. 3(a). The armchair nanoribbons of R-graphyne are uniformly metallic with zero band gaps except for the width $N_A$=1. As the width $N_A$=1, an indirect band gap with a value 0.25 eV appears, where the valence band maximum is located at X point and the conduction band minimum is lied at Γ point. The band gap of $N_A$=1 has a band gap should mainly come from the size effect, as found in graphyne nanoribbons[22]. Except for $N_A$=1, the band gap is always equal to zero regardless of the width, as the valence band crosses through Fermi level (See Fig. 3(b)). Such results indict that the armchair nanoribbon of R-graphyne remains the same electronic property as that of the 2-D R-graphyne.

Interestingly, unlike the armchair edged nanoribbons, the zigzag nanoribbons exhibit different electronic properties from the bulk graphyne. Below $N_Z$=2, it shows a larger band gap of 0.36 eV because of the quantum confinement effect. When $N_Z$>2, the band gaps of zigzag nanoribbons shows a remarkable half-integer oscillation behavior between semiconductor and metallic as a function of the width of nanoribbons. Recently, Luo et al[28] observed the oscillation between semiconductor and metallic for the bulk 2D super lattice, and they contribute the main reason from the symmetry. Here, the same kind of oscillation is observed in the nanoribbon instead of bulk. The nanoribbon with the integer width $N_Z$ is semiconductor with a finite band gap (about 0.25 eV), while the nanoribbon with half-integer width $N_Z$ exhibits the electronic properties of Dirac-like point with no band gaps at the Fermi level, which will be further identified in the following section. The oscillation behavior should mainly originate from structure symmetrical effect for the different width of nanoribbons.[13, 28] In detail, the structure of half-integer width owns central symmetry. Similarly, the oscillation behavior is also reported in nanoribbons of graphene-based zigzag type C4.[13] Different from those of graphyne nanoribbons, the electronic properties of planar C4 nanoribbons oscillate between semiconducting and semi-metallic or metallic.[13, 28] The graphyne nanoribbons are always semi-metallic properties regardless of the width.[22]

The detailed band structures of the zigzag nanoribbons are shown in Fig. 3(c). The typical nanoribbons with widths of $N_Z$=2.5, $N_Z$=3, and $N_Z$=3.5 are chosen to show the electronic properties of zigzag R-graphyne. As for the half-integer width (See $N_Z$=2.5 and $N_Z$=3.5 in Fig. 3 (c)), the valence band and conduction one meet at the Fermi level. Interestingly, the meeting point is located in the middle of Γ and X at the first Brillouin zone (BZ) similar to that of bulk buckled T-graphene or 6,6,12-graphyne.[9, 17] Such feature is different from graphene, which is at K high symmetry point. As for the half-integer number of repeating cell, except that the nanoribbon with width of $N_Z$=2.5 has only one meeting point, all others have two meeting points at Fermi level regardless of the widths of nanoribbons. The band structure of the nanoribbon with the integer number of width ($N_Z$=3) is a direct semiconductor with a band gap of 0.36 eV (See Fig. 3(c)). The valence band maximum and conduction band minimum not only locate at Γ point but also in the middle of Γ and X point in the first Brillouin zone.

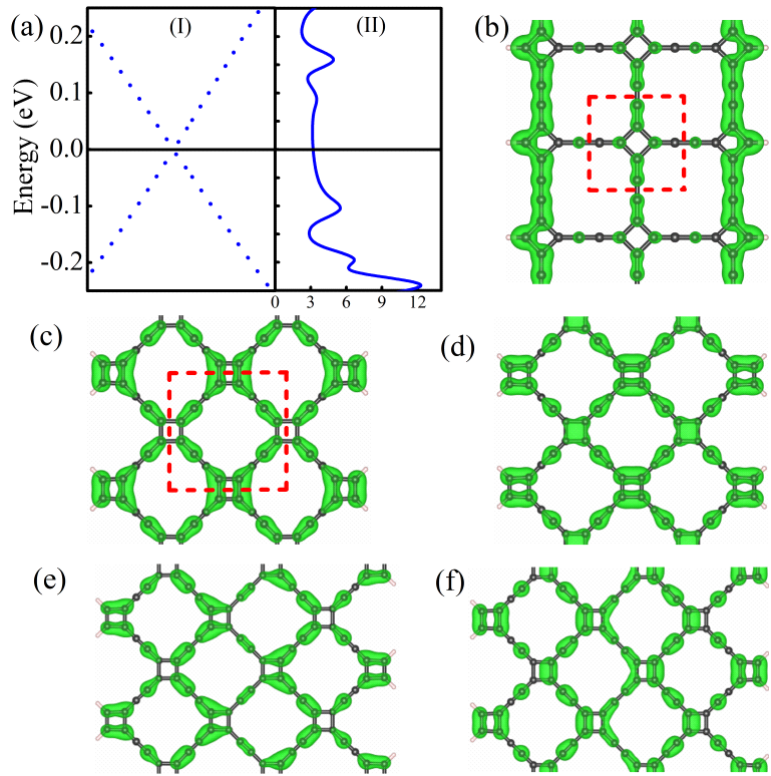

Fig. 4 (color online) The electronic structures and charge density distribution of R-graphyne nanoribbons. (a) Amplification of the band structure (I) and DOS (II) around the meeting point for $N_Z=2.5$. The charge density distribution near $E_F$ for armchair R-graphyne nanoribbons ($N_Z=3$) (b); The maximum point at valence band (c) / (e) and the minimum point at conduction band (d) / (f) for $N_Z=2.5/N_Z=3$. The black balls represent carbon atoms.

As discussed above, the half-integer number of repeating cell for zigzag nanoribbons have two meeting points at the Fermi level. To further check the electronic properties around the Fermi level, a typical zigzag ribbon with width of $N_Z=2.5$ is investigated, and the results are shown in Fig. 4(a [II]). It can be seen that the total DOS near the Fermi level is about 3 states/eV/cell, which is close to that of T-graphene with 1.04 states/eV/cell .[9] Such results suggest that the meeting point of zigzag nanoribbons with $N_Z=2.5$ is a Dirac-like one. The amplification band structure around the Fermi level (See Fig. 4(a [I])) exhibits a linear dispersion relation, and the calculated Fermi velocity is about $3.71 \times 10^5$ m/s based on the Fermi Dirac function $v_F \approx E(q)/q$. The Fermi velocity of graphene is about ~ $10^6$ m/s,[1,4] thus the Fermi velocity of half-integer zigzag R-graphyne is quite close to that of graphene, further validating that the meeting point is indeed a Dirac-like point.

In order to discover the physical origin of the Dirac-like points, the charge density distributions of R-graphyne nanoribbons are calculated near the $E_F$ for both kinds of edge styles (armchair and zigzag) with different widths, and the results are shown in Fig. 4(b)-4(f). The charge distribution of armchair nanoribbon with $N_A=3$ around $E_F$ (See Fig. 4(b)) clearly shows that the electrons are delocalized at the two edge carbon linkages, suggesting the metallic property, as shown in the band structure.

Unlike those of the armchair edged ones, the charge densities of the zigzag edged nanoribbons exhibit two different behaviors for half-integer and integer number of width, as observed in their band structures. As for the half-integer width (See Fig. 4(c), $N_Z=2.5$), the charge distribution at the maximum valence band shows that the electrons are not fully delocalized around the C4 ring. The orbitals of C4 overlaps with those of the neighboring atoms on the acetylenic linkages, but the orbitals of the middle C4 rings almost overlap to those of neighboring atoms on the C4 ring. As for the minimum conduction band, the orbitals are localized around the C4 and the acetylenic linkages, respectively, as shown in Fig. 4(d). The charge density distributions for both conduction band and valence band are in mirror symmetry along x- and y-directions. Meanwhile, the charge density emerges symmetry with respectively to the central line for the half-integer width. However, the charge distribution of integer width ($N_Z=3$) in the two sides does not have same kind of feature, which exhibits rotational symmetry. Thus the different structure symmetry make the different electronic distribution around the Fermi level, which should be the main reason for the different electronic structures: The half-inter width exhibits the intriguing Dirac-like point, while the integer width zigzag edged nanoribbons is semiconductor.

The different charge distribution between the half-integer and integer width zigzag nanoribbons suggests that the Dirac-like point is greatly affected by central symmetry rather than rotation symmetry. Graphene is hexagonal symmetry, 6,6,12-graphyne square symmetry, and buckled T-graphene also square symmetry with two comprising sub-lattices.[9] More importantly, the two type edge nanoribbons have different repeating unit cell. The armchair edged nanoribbons possess a C8 repeating unit cell, as shown in red square of Fig. 4(b). Such kind of repeating unit cell is similar to that of the bulk R-graphene (See Fig. 1(b). The zigzag edged nanoribbon has a C16 repeating unit cell in Fig. 4(c). Such repeating unit cell contains two sub-lattices. Recent study suggested that T-graphene containing two sub-lattice can exhibit the electronic property of Dirac-like point[9]. Considering the zigzag type nanoribbons contain two sub-lattices, it should be another important origin for the emergence of Dirac-like fermions.

## IV. Conclusions

In summary, R-graphyne is demonstrated to be a novel 2D carbon allotrope, which is thermodynamically stable. The electronic structures show that the R-graphyne is metallic as the valence band cross the Fermi level. Interestingly, the R-graphene exhibits intriguing electronic properties through tailoring of the bulk material in different directions and width. Although the armchair stripe is still metal with no band gaps, while the zigzag R-graphyne exhibits extraordinary half-integer oscillation behavior between metal and semiconductor. Of particularly interesting, the zigzag R-graphyne nanoribbons with half-integer widths show Dirac-like fermions with a Fermi velocity of $3.7 \times 10^5$ m/s. The presence of Dirac-like fermions in the nanoribbons is orientation-dependent. The further analysis of electronic structure unveils that the Dirac-like fermions of zigzag nanoribbons origins from both two sub-lattices and central symmetry. The unique electronic properties of R-graphyne nanoribbons greatly enrich our understanding on origin of Dirac-like clone, and such material has great potential for nanosize electronic applications.

**Acknowledgments**

This work was supported by the National Natural Science Foundation of China (Nos. 51006086, 11074213, 51176161, 11244001, and 51032002), the CAEP foundation (Grant No. 2012B0302052), the MOST of China (973 Project, Grant NO. 2011CB922200). The computations supports from Informalization Construction Project of Chinese Academy of Sciences during the 11th Five-Year Plan Period (No.INFO-115-B01) are also highly acknowledged.